\documentstyle [12pt]{article}
\makeatletter \oddsidemargin  0in \evensidemargin 0in
\textwidth16cm \RequirePackage[dvips]{graphicx} \textheight 20.5cm
\newcommand{\singlespacing}{\let\CS=\@currsize\renewcommand{\baselinestretch}{1.5}\tiny\CS}
\makeatletter \oddsidemargin  -.1in \evensidemargin -.1in
\newcommand{\doublespacing}{\let\CS=\@currsize\renewcommand{\baselinestretch}{1.35}\tiny\CS}

\def\@citex[#1]#2{\if@filesw\immediate\write\@auxout{\string\citation{#2}}\fi
  \def\@citea{}\@cite{\@for\@citeb:=#2\do
    {\@citea\def\@citea{,\linebreak[0]\hskip0pt plus .2em}%
      \@ifundefined{b@\@citeb}%
    {{\bf ?}\@warning{Citation `\@citeb' on page \thepage\space undefined}}%
      \hbox{\csname b@\@citeb\endcsname}}}{#1}}

\newtheorem{rule-def}[theorem]{Rule}
\begin{document}
\title{\bf Non Existence of Quantum Mechanical Self Replicating Machine}\author{I. Chakrabarty$^{a}$, Prashant$^{b}$
\thanks{Corresponding author: Prashant, email: prashant.iiitm@gmail.com}\\
$^{a}$Heritage Institute of Technology, Kolkata-107, West Bengal, India\\
$^{b}$ Indian Institute of Information Technology and Management,
India. }
\date{}
\maketitle{}
\begin{abstract}
In this paper we establish the impossibility of existence of self
replicating machine in the quantum world. We establish this result
by three different but consistent ways of linearity of quantum
mechanics, no signalling criterion and conservation of entanglement
under Local Operation and Classical Communication.
\end{abstract}
\section{\bf Introduction:}
It was Shannon who first introduced us to the amazing world of
binary logic and information theory. After that it had been a long
journey and still we are searching for the nature of information in
quantum world. The most important query from the physicists point of
view is to find out the answer to the question that how the laws of
physics are governing the dynamics of living units? Is it classical
physics which explains all the active forces of life or we have to
search into the microscopic world so that quantum mechanics can give
a valid explanation of it. One of the major thrust area of research
is to look out how the information defined in classical world
differs from quantum information and also to enlist many
computational procedures which are feasible in the classical world
but get restricted at microscopic level. Cloning and deletion which
is feasible in classical information, cannot be executed with
hundred percent fidelity for an unknown quantum state [1,2].
Recently it had been seen that self replication machine which is
feasible in classical information theory [3] fails to self replicate
two non orthogonal quantum states [4]. It may be remarked that the
process of replication or cloning and self replication is not just
the same thing. A Universal constructor is said to exist if it can
implement copying the original state along with the stored programme
by a linear operator
acting on the joint Hilbert Space.\\
There are many physicists [5] and biologists [6] who strongly
believe that quantum mechanics will solve the mysteries of living
systems.It had been suggested that the working principles of enzymes
could be understood from the quantum mechanical principles
[7].Quantum mechanical algorithms had been proposed for genetic
evolution [8].  It had been seen that quantum mechanics play an
important role during cell mutation and entanglement between the
mutational state and environment can also enhance the probability
of mutation [9].\\
In this work our basic objective is to prove the impossibility of
the self replication process in quantum world from linearity of
quantum mechanics, no signalling criterion and principle of
conservation of entanglement. Moreover here we try to give a
biological outlook of the self replication process. We also try to
find out what meaning does it convey to us biologically when we say
that self replication is not feasible because of the linear
structure of quantum theory,  no signalling condition and
conservation of entanglement under LOCC operations.
\section{\bf Non
existence of Self replicating machine: Linearity of Quantum
mechanics}A Quantum mechanical self replicating machine may be
completely specified by a quadruple
$(|\psi\rangle,|P_{U}\rangle,|C\rangle,|\Sigma\rangle)$. Here
$|\psi\rangle \in H^{N}$ is the state of the (artificial or real)
living system that contains quantum information to be self
replicated and $|P_{U}\rangle \in H^{K}$ is the programme state
that carries the instructions to copy the original information
(the unitary operator $U$;
$U(|\psi\rangle|0\rangle)=|\psi\rangle|\psi\rangle$, is encoded in
the programme state $|P_{U}\rangle$). Here $|C\rangle$ is the
control unit and $|\Sigma\rangle$ is the finite collection of the
blank states $|0\rangle|0\rangle|0\rangle.....|0\rangle$ in a $M$
dimensional Hilbert space on which the original state along with
the program state is to be copied. We had already seen in [4] such
a self replicating machine don't exist in quantum world and this
result is consistent with the unitarity of quantum theory.\\
In this section we try to find out whether the linear structure of
Quantum theory supports the existence of self replicating machine.
Here we will treat quantum states as a smallest state in the
living system that represents either
'artificial' or 'real' life.In this process we will assume a self replicating machine for orthogonal quantum
states with non orthogonal program states.Then we will try to see that whether we can self replicate an unknown qubit. \\
Let $|\psi_1\rangle$ and $|\psi_2\rangle$ are two orthogonal
quantum states and let $|P_{U_1}\rangle$ and $|P_{U_2}\rangle$ be
the programme states, where the unitary operators for copying
those two orthogonal states are encoded. Here $|P_{U_1}\rangle$
and $|P_{U_2}\rangle$ are non orthogonal quantum states as we know
that Quantum mechanical self replicating machine for orthogonal
states can exist only when the program states are non
orthogonal[15]. Therefore the self replicating process for these
two orthogonal quantum states is given by,
\begin{eqnarray}
L[|\psi_1\rangle|0\rangle|P_{U_1}\rangle|0\rangle^{\otimes
m}|C\rangle]|0\rangle^{\otimes
n-(m+1)}=|\psi_1\rangle|P_{U_1}\rangle
L[|\psi_1\rangle|0\rangle|P_{U_1}\rangle|0\rangle^{\otimes
m}|C^1\rangle]|0\rangle^{\otimes n-2(m+1)}\\
L[|\psi_2\rangle|0\rangle|P_{U_2}\rangle|0\rangle^{\otimes
m}|C\rangle]|0\rangle^{\otimes
n-(m+1)}=|\psi_2\rangle|P_{U_2}\rangle
L[|\psi_2\rangle|0\rangle|P_{U_2}\rangle|0\rangle^{\otimes
m}|C^2\rangle]|0\rangle^{\otimes n-2(m+1)}
\end{eqnarray}
It is important that (1-2) is not merely a cloning transformation,
on the contrary it is a recursively defined transformation where
the fixed unitary operator L acts on initial (parent)
configuration and the same operator acts on the final child
configuration after the copies have been produced.\\
Let us consider a non orthogonal quantum state $|\xi\rangle$ which
can be expressed as a linear superposition of orthogonal quantum
states $|\psi_1\rangle$ and $|\psi_2\rangle$.\\Let
\begin{eqnarray}
|\xi\rangle=\alpha|\psi_1\rangle+\beta|\psi_2\rangle
\end{eqnarray}
where $\alpha^2+|\beta|^2=1$.\\ Let us consider the action of a
quantum mechanical self replicating machine  on an unknown quantum
state $|\xi\rangle$ along with the programmed state
$|P_{U}\rangle$ on basis of the transformation defined in (1-2)
\begin{eqnarray}
&&L[(\alpha|\psi_1\rangle|0\rangle(|P_{U_1}\rangle)+\beta|\psi_2\rangle|0\rangle(|P_{U_2}\rangle))|0\rangle^{\otimes
m}|C\rangle]|0\rangle^{\otimes n-(m+1)}={}\nonumber\\&&\alpha
|\psi_1\rangle|P_{U_1}\rangle
L[|\psi_1\rangle|0\rangle|P_{U_1}\rangle|0\rangle^{\otimes
m}|C^1\rangle]|0\rangle^{\otimes n-2(m+1)}+\beta
{}\nonumber\\&&|\psi_2\rangle|P_{U_2}\rangle
L[|\psi_2\rangle|0\rangle|P_{U_2}\rangle|0\rangle^{\otimes
m}|C^2\rangle]|0\rangle^{\otimes n-2(m+1)}
\end{eqnarray}
which is not equivalent to the original self replicating process
defined as
\begin{eqnarray}
&&L[|\xi\rangle|0\rangle|P_{U}\rangle|0\rangle^{\otimes
m}|C\rangle]|0\rangle^{\otimes n-(m+1)}=|\xi\rangle|P_{U}\rangle
L[|\xi\rangle|0\rangle|P_{U}\rangle|0\rangle^{\otimes
m}|C^3\rangle]|0\rangle^{\otimes n-2(m+1)}
\end{eqnarray}
Since (4)and (5)can never be identical on basis of the linear
structure of quantum theory.Thus we can say that construction of
such a machine in quantum world for an unknown qubit is strictly
impossible.In real biological systems self replication of a
macroscopic species is a classical process that takes place in an
open system in which decoherence is very strong and rapid. This
can be probably one explanation of the feasibility of self
replication process in biological world. Self replication process
at the quantum level may resemble biological replication while in
reality replication of a living organism may be something
different and should not be confused with the former one.
\section{\bf Non
existence of Self replicating machine:No signalling principle and
Conservation of entanglement under LOCC } In this section we try
to find out whether the fundamental principles like no-signalling criterion
and conservation of entanglement under LOCC support the existence
of quantum mechanical universal self replicating machine.\\
Let us consider two non orthogonal states given by the form,
\begin{eqnarray}
|\psi_1\rangle=a|0\rangle+b|1\rangle,\\
|\psi_2\rangle=c|0\rangle+d\exp{i\theta}|1\rangle
\end{eqnarray}
where~$ a,b,c,d$ are real numbers satisfying the
relation~$a^2+b^2=c^2+d^2=1$~and~$0<\theta<\pi,a>0,c>0$ and the
states $|0\rangle $ and $|1\rangle $ are orthogonal to each
other.Let $|P_{U_1}\rangle$ and $|P_{U_2}\rangle$ are the
programme states
corresponding to the states $|\psi_1\rangle$ and $|\psi_2\rangle$ respectively.\\
Let us consider an entangled state shared by two distant parties
Alice and Bob
\begin{eqnarray}
|\chi\rangle&=&\frac{1}{\sqrt{2}}|0\rangle_A[|\psi_1\rangle|\Sigma\rangle|P_{U_1}\rangle|\Sigma\rangle^{\otimes
m}|C\rangle]_B(|\Sigma\rangle^{\otimes
n-(m+1)})_B+{}\nonumber\\&&\frac{1}{\sqrt{2}}|1\rangle_A[|\psi_2\rangle|\Sigma\rangle|P_{U_2}\rangle|\Sigma\rangle^{\otimes
m}|C\rangle]_B(|\Sigma\rangle^{\otimes n-(m+1)})_B
\end{eqnarray}
where Alice is in possession with the qubit 'A'and Bob is in
possession with the qubit 'B'.\\
Let us assume that Bob is in possession with quantum mechanical
universal constructor whose action on the non orthogonal quantum
states is defined by the transformations,
\begin{eqnarray}
&&L([|\psi_1\rangle|\Sigma\rangle|P_{U_1}\rangle|\Sigma\rangle^{\otimes
m}|C\rangle](|\Sigma\rangle^{\otimes
n-(m+1)}))=|\psi_1\rangle|P_{U_1}\rangle
L([|\psi_1\rangle|\Sigma\rangle|P_{U_1}\rangle|\Sigma\rangle^{\otimes
m}|C^1\rangle]{}\nonumber\\&&
(|\Sigma\rangle^{\otimes n-2(m+1)}))\\
&&L([|\psi_2\rangle|\Sigma\rangle|P_{U_2}\rangle|\Sigma\rangle^{\otimes
m}|C\rangle](|\Sigma\rangle^{\otimes
n-(m+1)}))=|\psi_2\rangle|P_{U_2}\rangle
L([|\psi_2\rangle|\Sigma\rangle|P_{U_2}\rangle|\Sigma\rangle^{\otimes
m}|C^2\rangle]{}\nonumber\\&& (|\Sigma\rangle^{\otimes n-2(m+1)}))
\end{eqnarray}
The reduced density matrix on Alice's side before the application
of quantum mechanical universal constructor is given by,
\begin{eqnarray}
\rho^A=\frac{1}{2}[I+|0\rangle\langle1|(\langle\psi_2|\psi_1\rangle\langle
P_{U_2}|P_{U_1}\rangle)+
|1\rangle\langle0|(\langle\psi_1|\psi_2\rangle\langle
P_{U_1}|P_{U_2}\rangle)]
\end{eqnarray}
The reduced density matrix on Alice's side after the application
of quantum mechanical universal constructor is given by
\begin{eqnarray}
\rho^A_U=\frac{1}{2}[I+|0\rangle\langle1|(\langle\psi_2|\psi_1\rangle^2\langle
P_{U_2}|P_{U_1}\rangle^2\langle C^2|C^1\rangle)+
|1\rangle\langle0|(\langle\psi_1|\psi_2\rangle^2\langle
P_{U_1}|P_{U_2}\rangle^2\langle C^1|C^2\rangle)]
\end{eqnarray}
Since this operation is totally local and there is no classical
communication between two parties ,the density matrix on Alice's
side must remain unchanged .Now from equations (11) and (12) we
get that the equations will be identical only when
\begin{eqnarray}
\langle\psi_2|\psi_1\rangle\langle
P_{U_2}|P_{U_1}\rangle[1-\langle\psi_2|\psi_1\rangle\langle
P_{U_2}|P_{U_1}\rangle\langle C^2|C^1\rangle]=0
\end{eqnarray}
or
\begin{eqnarray}
\langle\psi_1|\psi_2\rangle\langle
P_{U_1}|P_{U_2}\rangle[1-\langle\psi_1|\psi_2\rangle\langle
P_{U_1}|P_{U_2}\rangle\langle C^1|C^2\rangle]=0
\end{eqnarray}
Now the equation (13) tells us that the Self replicating machine
exists under two criterions, either i)
$\langle\psi_2|\psi_1\rangle=0$, $\langle
P_{U_2}|P_{U_1}\rangle\neq0$. or ii)
$\langle\psi_2|\psi_1\rangle\neq0$, $\langle
P_{U_2}|P_{U_1}\rangle=0$.\\
The first condition states that if the states are orthogonal, no
restrictions are imposed on the program state. This clearly
indicates that with finite dimensional program state and finite
number of blank states orthogonal states can self replicate.This
is nothing but a realization of classical universal constructor
[3]. On the other hand we see that the non orthogonal quantum
states can self replicate only when the program states are
mutually orthogonal in a finite dimensional program Hilbert space.
Since here the self replication process is totally a local
operation and there is no classical communication from Bob to
Alice, the reduced density matrix on Alice's side will remain
unchanged from the principle of no signalling. But we find that
the equations (11) and (12) will not be identical unless either of
the conditions (i) and (ii) holds. Hence we come to the conclusion
that the transformations defined in (9,10) is not valid. This
establishes the impossibility of the existence of quantum
mechanical universal constructor from the principle of no
signalling. In living organisms the concept of entanglement is
nothing but interdependency between different species. The
probable biological explanation to the principle of no signalling
is that in absence of classical communication between two species,
any local operation like self replication on one of the correlated
species doesn't change the state of
the other organisms.\\\\
Let us consider the largest eigen values $\lambda^A$ and
$\lambda^A_U$ of the two density matrices (11) and (12)
respectively
\begin{eqnarray}
\lambda^A=\frac{1}{2}+\frac{|p||q|}{2}\\
\lambda^A_U=\frac{1}{2}+\frac{|p|^2|q|^2|r|}{2}
\end{eqnarray}
where,\\ $p=\langle\psi_1|\psi_2\rangle\\
q=\langle P_{U_1}|P_{U_2}\rangle\\
r=\langle C^1|C^2\rangle$\\
Now $\lambda^A-\lambda^A_U=\frac{1}{2}|p||q|[1-|p||q||r|]$.Since
$|p|<1$,$|q|<1$ and $|r|<1$,therefore $|p||q||r|\neq1$.Hence
$\lambda^A-\lambda^A_U\neq0$.\\
Now, $\lambda^A\neq\lambda^A_U$ and hence the amount of
entanglement of the entangled state before and after the
application of self replicating machine doesn't remains same. This
violates the conservation of entanglement under LOCC. The
violation of conservation of entanglement is equivalent of saying
the violation of the principle of conservation of information in
the closed system [13,14]. Hence once again we rule out the
existence of universal self replicating machine from the principle
of conservation of entanglement.
\section{\bf Conclusion:}In this paper we have considered an artificial living system, where quantum states are
treated as living units carrying information. We have then
analyzed the process of self replication as a microscopic
phenomenon. We came to a conclusion that the process of self
replication is not possible for quantum states, by we assuming
either the linear structure of quantum theory, the principle of no
signalling or conservation of entanglement to be valid. In
reality, we know that the process of self replication is feasible
in the biological world even if the laws of quantum mechanics
prevail at the level of living organisms[11]. This may indicate
there is something beyond these principles that may be the root.
Thus by self replication of a living unit we never refer to a
situation which is going to violate these principles.Indeed it
remains a major question that how this restriction will look in
light of no signalling and conservation of information. If we go
for the explanation of biological processes like reproduction of a
living unit from information theoretic viewpoint we see that the
information obtained by the living unit to carry out the process
of self replication already existed somewhere; thus by
establishing the permanence of information [5,12].
\section{\bf Acknowledgement:}
Indranil Chakrabarty acknowledges Prof B.S.Choudhury, S. Adhikari,
Prof. C.G.Chakraborti for their encouragement and inspiration in
completing this work.  Both authors acknowledge Prof A.K.Pati, for
providing encouragement and support in completing this work.
\section{\bf Reference:}
$[1]$ W.K.Wootters and W.H.Zurek,Nature 299 (1982) 802-803.\\
$[2]$ A.K.Pati and S.L.Braunstein Nature 404 (2000) 164.\\
$[3]$ J.Von Neumann, The theory of Self-Replicating Automata.
University of Illinois Press,Urbana,IL (1966)(work by J.von
Neumann in 1952-53).\\
$[4]$ A.K.Pati and S.L.Braunstein, Quantum mechanical universal
constructor,quantph /0303124 (2003).\\
$[5]$ E.Schrodinger,What is life?, Cambridge University
Press,London,(1944).\\
$[6]$ R.Penrose, Shadows of Mind, Oxford University
Press,Oxford,(1944).\\
$[7]$ J.McFadden, Quantum Evolution, Harper Collins, London
(2000).\\
$[8]$ H.Frolich, Nature 228 (1970) 1093.\\
$[9]$ A.Patel,J.of Bioscience 26 (2001) 145-151.\\
$[10]$ J.McFadden and J.Al-Khalili, Biosystems 50(1999) 203-211.\\
$[11]$ A.K.Pati, Replication and Evolution of Quantum
Species,quant-ph/ 0411075.\\
$[12]$ R.Jozsa, A stronger no-cloning theorem. quant-ph/0204153
(2002).\\
$[13]$ M.Horodecki.et.al,No-deleting and no-cloning principles as
consequences of conservation of quantum in
formation,quant-ph/0306044.\\
$[14]$ M.Horodecki.et.al,Common origin of no-cloning and
no-deleting - Conservation of information,quant-ph/0407038.\\
$[15]$ M. A. Nielsen, I. Chuang, Phys. Rev. Lett. 79, 321-324
(1997).
\end{document}